\documentclass[12pt,preprint]{aastex}
\received{}
\revised{}
\accepted{}

\shorttitle{3C 273}

\begin{document}
\title{{\sl Chandra} Detection of Local \ion{O}{7} He$\alpha$
Absorption Along the Sight Line Toward 3C~273}

\author{Taotao~Fang\altaffilmark{1,2},
	Kenneth~R.~Sembach\altaffilmark{3}, AND
	Claude~R.~Canizares\altaffilmark{1}}
	\altaffiltext{1}{Department of Physics and Center for Space
	Research, MIT,  77 Mass. Ave., Cambridge, MA 02139, }
	\altaffiltext{2}{Department of Physics, Carnegie Mellon University, 
	5000 Forbes Ave., Pittsburgh, PA 15213} 
	\altaffiltext{3}{Space Telescope Science
	Institute, 3700 San Martin Dr.,  Baltimore, MD 21218}

\begin{abstract}

With the {\sl Chandra} X-ray Telescope we have detected a zero-redshift
\ion{O}{7}
He$\alpha$  absorption line along the sight line toward 3C~273. 
This line detection is  highly significant, with a signal-to-noise
ratio (S/N) of 6.4. We explore two  models, which associate this line
with (1) the intragroup gas in the Local Group, and (2) the hot halo gas in the vicinity of our Milky Way. In  the first model,
we find that for a standard $\beta$-model of the gas distribution  in the
Local Group, the temperature is constrained to $\rm 2.3\times 10^5 < T < 1.2\times 10^6$ K and the baryon overdensity is $\delta_b \sim 100$; both results are
consistent with the properties of the so-called warm-hot intergalactic
medium (WHIM) predicted by cosmological simulations. We also find the
core radius of the gas distribution should be $> 100$ kpc. In the
second model we discuss several possible Galactic origins for the 
absorption, and we comment on the possibility that the \ion{O}{7} is
associated with the  \ion{O}{6} absorption observed in this direction by
the Far Ultraviolet Spectroscopic Explorer ($\sl FUSE$). We find that there is 
a strong indication that collisional ionization is the dominant ionization 
source for the observed absorption.

\end{abstract}

\section{Introduction}

Highly ionized resonance absorption lines from heavy elements (such
as \ion{C}{4}, \ion{Si}{4}, \ion{N}{5}, \ion{O}{6}) at low redshift
have been detected along many lines of sight by the Far
Ultraviolet Spectroscopic Explorer ($\sl FUSE$)  and the Hubble Space
Telescope ($\sl HST$) (e.g., \citealp{tsj00,sem00,sav00}). These
species arise in the interstellar medium of the Milky Way and the 
low redshift intergalactic medium (IGM).  Among these ions, \ion{O}{6} is 
of particular interest because it is an excellent tracer of gas at 
 $T \geq 10^5$ K. The detection of
\ion{O}{6} in many directions (\citealp{sav02}; \citealp{sem02}; 
\citealp{wak02}) indicates the existence of hot gas in the halo of the 
Milky Way, in high velocity clouds surrounding the Milky Way, or even in the Local Group \citep{nze02}.

Recently, with the unprecedented high resolution of the {\sl Chandra}
X-ray Telescope, \citet{nic02} and \citet{fan02}
reported the first detection of X-ray absorption lines from higher ionization
species, \ion{O}{7} and
\ion{O}{8}, at $z\approx 0$ along the sight line toward the quasar 
PKS~2155-304. \citet{mwc02} reported possible detection of such X-ray absorption feature in the spectrum of H~1821+643.  \citet{kas02} also
reported the detection of zero-redshift \ion{O}{7} in a long exposure 
($\sim$ 900 ksec) of the bright
Seyfert galaxy NGC~3783 with {\sl Chandra}. Both \ion{O}{7} and 
\ion{O}{8} trace
gases with temperatures  $\gtrsim10^6$\,K. Therefore, it is of
great interest to see whether the higher temperature gas is related
to the gas that produces the UV absorption lines and whether there is 
a multi-phase distribution of hot gas in the vicinity of the Milky
Way. It is also possible that the X-ray absorption is associated
with hot intragroup gas in the Local Group. If so, this might
provide important information about the ``missing baryons'' at low 
redshift.  Cosmological simulations indicate that a large fraction of 
neutral gas at high redshift is heated to high temperatures 
($ 10^5 < T < 10^7$ K) at $z\sim 0$
(\citealp{cko95,oce96,fhp98,cos99}); gas in the higher end of this 
temperature range  has so far escaped detection 
 at low redshift. Simulations predict that these ``missing baryons''
 are distributed in intergalactic 
space in the form of diffused filamentary structures (the  warm-hot intergalactic medium, or WHIM) that connect virialized clusters and/or groups of 
galaxies with ovdensities of $\delta_b \sim 5 - 200$.

In this {\sl Letter} we report the detection of a $z\sim0$ \ion{O}{7}
He$\alpha$ resonance absorption line in the {\sl Chandra} spectra of
3C~273.  We explore two  possible sources of this absorption:
 (1) intragroup gas in the Local Group, and (2) hot gas in the 
 Milky Way halo and interstellar medium along the sight line. 

\section{Data Analysis}

3C~273 ($l=289.95^{\circ}$, $b=64.36^{\circ}$, $z_{QSO}=0.158$) is one
of the brightest quasars in the sky at X-ray wavelengths 
and has been repeatedly observed as a
{\sl Chandra} calibration target. It was observed with the {\sl
Chandra} Low Energy Transmission Grating Spectrometer (LETGS) and the 
Advanced CCD Imaging Spectrometer (ACIS)
 on 31 October 2000 (obs. \#1198), 13 June 2001 (obs \#2464), and 15 June 2001
(obs \#2471), with an exposure time of 40, 30 and 30 ksec, respectively. See the {\sl Chandra} Proposers' Observatory Guide at
http://asc.harvard.edu/ for more information about {\sl Chandra} and 
its instrumentation. In the analysis and discussions that follow, we adopt 
$\rm H_0 =  70\,h_{70}~ km~s^{-1}~Mpc^{-1}$. Unless
otherwise stated, errors are quoted at 90\% confidence.

We reduced the Level 1 data using standard CIAO tools\footnotemark
\footnotetext{Chandra Interactive Analysis of Observations (CIAO)", see http://asc.harvard.edu/ciao/}, and 
restricted all spectra to the 2--42\,\AA\ wavelength range to avoid
complexities at longer wavelengths from the instrumental Carbon K edge.
We summed the events in 0.025\,\AA\ bins, which are half the width of 
the instrumental line response function (FWHM $\sim
0.05$\,\AA\ across the bandpass).
To remove the underlying broad-band continuum
behavior for each dataset, we divided each spectrum by the best fit 
 power-law, with a correction for absorption by the foreground Galactic 
interstellar medium.
 We then fit the remaining residuals  with a sixth-order
polynomial. These procedures remove spectral features larger than
$\sim 8$\,\AA\ in width (e.g., broad 5-10\%  calibration uncertainties) but
preserve narrow line features. We combined the data from the three observations
as a weighted sum accounting for differences in exposure times 
and effective areas.

All of the continua are  well described by a single power law with a 
similar power-law index. The power-law
photon  indices are $\Gamma_{1198} \sim 2.08 \pm 0.36, \Gamma_{2464}\sim
1.86 \pm  0.40$,\ and\ $\Gamma_{2471}\sim 1.82 \pm 0.48$, assuming 
absorption by a Galactic neutral hydrogen column of 
$N_{H\,I} = 9\times 10^{19}~{\rm cm^{-2}}$ (Lockman \&
Savage~1995). The October 2000 observation exhibits the strongest
soft X-ray flux; $F_{\rm 0.1-2.4\,keV} = (9.3, 7.9, 7.8 )\times 10^{-11}~{\rm
ergs~cm^{-2}s^{-1}}$ for observations \#1198, \#2464, and \#2471, respectively. The total counts per bin in the continuum decline smoothly from $\sim$ 150 at 18\,\AA\ to $\sim$ 40 at 23\,\AA, with $\sim$ 70 at 21.6\,\AA.

After a blind search for any statistically significant absorption
features in the 2--42\,\AA\ region of the LETGS spectral bandpass
(Figure~1), we identified an absorption feature with S/N~$\sim 6.4$ at
$21.60\pm0.01$\AA, which is almost identical to the rest wavelength of \ion{O}{7} He$\alpha$ (see Table~1). We fit this feature  with single-component
Gaussian models using ISIS\footnotemark, taking into account the
instrumental line response function (Table~1).
\footnotetext{Interactive Spectral Interpretation System, see \citealp{hde00}.}
 We also examined other {\sl
Chandra} LETGS archival data for 3C\,273 and found a similar feature 
at the same wavelength with a different focal plane detector - the
High Resolution Camera (HRC).  

\section{Local \ion{O}{7} He$\alpha$ Absorption}

Based on the detected line equivalent width ($W_{\lambda}$), we can
estimate  the implied column density of \ion{O}{7}
\citep{spi78}. Since the  line is unresolved, the Doppler
$b$-parameter must be $b=\sqrt{2}\sigma \rm < 400~km~s^{-1}$ (90\%
confidence). If $b > 200\rm~km~s^{-1}$, then the line is unsaturated
and N(O~{\sc vii})$\rm = 1.8_{-0.7}^{+2.7} \times 10^{16}~cm^{-2}$.
The line becomes saturated at $b \sim 100\rm~km~s^{-1}$, giving
N(O~{\sc vii})$\rm \sim 10^{17}~cm^{-2}$. At $b < 100\rm~km~s^{-1}$ the
line is heavily saturated, but this can be ruled out by the lack of
higher Lyman series lines in our spectrum. From the non-detection of
local \ion{O}{7} He$\beta$ line at $18.6288\AA$, we estimate a
$4\sigma$ equivalent width upper limit of 11.9$\rm~m\AA$. Based
on a curve-of-growth calculation for \ion{O}{7} He$\beta$, we find that 
even in the case for which the line is heavily saturated 
($b < 100\rm~km~s^{-1}$),
the He$\beta$ upper limit restricts the \ion{O}{7} column density to 
$< 2\times10^{16}~\rm cm^{-2}$.  We therefore 
adopt N(O~{\sc vii})$\rm=1.8_{-0.7}^{+0.2}
\times 10^{16}~cm^{-2}$ in the following discussion.

Highly ionized metals such as \ion{C}{4} and \ion{O}{6} at $z\approx
0$ have been detected repeatedly through their absorption-line
signatures along many lines of sight (see \citealp{sav02}, 
\citealp{sem02}, and references therein).   
Unlike these lower ionization species, which typically trace transition
temperature gas with $T\sim (1 - 5) \times 10^{5}$ K, the \ion{O}{7}
detected with {\sl Chandra} indicates the existence of the gas with
temperatures $T\sim 10^{6}$ K. This high temperature gas traced by \ion{O}{7}
is unlikely to be produced in the nearby interstellar medium (ISM), 
although we cannot rule out the possibility of a nearby supernova remnant
origin. It is more
plausible that the He-like Oxygen is produced in hot Galactic halo
gas, or even in the Local Group. It is to these two possibilities that
the rest of this ${\sl Letter}$ is devoted.

\subsection{Comments on a Possible Local Group Origin}

The detection of high column density \ion{O}{7} implies the existence
of a large amount of hot gas with a temperature around $10^{6}$ K. In the
case of pure collisional ionization, temperature is the only parameter
of importance over a wide range of density so long as the gas is
optically thin.  The \ion{O}{7} ionization fraction peaks at nearly 1,
and exceeds 0.5, for temperatures $T \sim (0.5-1.3) \times10^{6}$~K
\citep{sdo93}. 
However, the IGM is also photoionized by the background UV/X-ray
radiation, which increases the ionization fraction at low temperatures
and densities. We explored these regimes using a grid of CLOUDY models
\citep{fer98} for $10^{2} < T < 10^{9}$ K and $10^{-8} < n ({\rm
cm^{-3}}) < 10^{-2}$ with 0.2$Z_{\odot}$. We used the UV background
from \citet{shu99}, and a power law with an exponential cutoff
at 50 keV and $\alpha=0.29$ in the X-ray spectral region 
(see, e.g. \citealp{fba92}). We combined the ionizing UV and X-ray spectra 
using
normalizations of $J_{13.6\,{\rm eV}} = 1.8\times10^{-23}\ {\rm ergs\
cm^{-2}~s^{-1}~Hz^{-1}~sr^{-1}}$ and $J_{1\,{\rm keV}} = 10\ {\rm ph\
cm^{-2}~s^{-1}~keV^{-1}~sr^{-1}}$. CLOUDY calculations indicate that the
ionization fraction versus $T$ relation begins to deviate appreciably from
the pure collisional case for low density regions. At
a baryon density $n_{b} \leq 10^{-4}\rm~ cm^{-3}$,  photoionized low temperature
gas ($ 4 \leq \log T \leq 5.5$) can produce significant amount of
\ion{O}{7}. We also considered several other background radiation models,
such as the X-ray background from \citet{miy98} and the UV background from
\citet{hma96},  but as shown  by \citet{che02} we found that 
the  resulting differences in the predicted \ion{O}{7} ionization fraction 
are negligible.

The upper limit on  the line width of $\rm \sigma \sim 400~km~s^{-1}$
sets an maximum to the path length of $\sim
5$\,h$^{-1}_{70}$\,Mpc, since otherwise the differential Hubble flow
would excessively broaden the line. We can further constrain the
density of the absorbing gas, assuming it is associated with the Local
Group. The path length can be set equal to the distance to the
boundary of the Local Group, where the gas begins to participate in
the Hubble flow. Assuming a simple spherical geometry for the Local
Group, the path  length is $\sim 1$ Mpc (see the following
text). Adopting the 90\% lower limit of the \ion{O}{7} column density,
this gives $n_{b} > (2.2\times10^{-5}\ {\rm cm}^{-3})\
Z_{0.2}^{-1}f_{1}^{-1}l_{1}^{-1}$, where $Z_{0.2}$ is the metallicity
in units of 0.2 solar abundance, $f_{1}$ is the ionization fraction,
and $l_{1}$ is the path length in units of 1 Mpc. This implies a
baryon overdensity of $\delta_{b} \sim$ 100, which is roughly
consistent with the overdensity of the WHIM gas predicted from
cosmological simulations (\citealp{cko95,oce96,cos99,dav01}).

A model characterizing the distribution of the Local Group gas is given by the standard
$\beta$-model \citep{cfu76}:
\begin{equation}
n(r) =
n_{0}\left[1+\left(\frac{r}{r_{c}}\right)^2\right]^{-\frac{3}{2}\beta}
\end{equation}
where $n_{0}$ is the central gas number density, $r_{c}$ is the core
 radius, $\beta$ is the ratio of the specific kinetic energy to
 thermal energy used to characterize the depth of the potential
 well. Poor groups like the Local Group typically have shallow
 potentials with $\beta \sim 0.5$ \citep{mdm96}, so we adopt
 this value. We adopt a simplified geometry model of the Local Group, 
 where the barycenter is located along the line connecting M31 and the Milky
 Way in the direction towards $(l,b) \approx (121.7^{\circ},
 -21.3^{\circ})$ at a distance of about 450 kpc from the Milky Way 
 \citep{rpe01}. 
 At about $r \sim 1200$ kpc, the gravitational
 contraction of the Local Group starts to dominate the Hubble flow, so this 
 is typically defined as the boundary of the Local Group \citep{cva99}.

Based on this simple model, we estimate the column density of
\ion{O}{7} along the 3C~273  sight line through the Local Group
by integrating the \ion{O}{7} number density $n_{O\,VII}$ out to the 
boundary.  Figure~2 shows the \ion{O}{7} column density ($\rm N_{O\,VII}$) 
versus baryon density at the center of the Local Group
($n_{0}$). By
applying a conservative upper limit of the Local Group gravitational
mass and arguing that the cooling timescale should be longer than half the
Hubble time, \citet{rpe01} obtained an upper limit of
the central density $n_{0} \leq 5\times10^{-4}\ {\rm cm^{-3}}$.
From Figure~2 we find a tight upper limit of
temperature $T \leq 1.2\times10^6$ K; at temperatures higher than $1.2\times10^6$ K, the
\ion{O}{7} ionization fraction drops quickly and \ion{O}{8} starts to
dominate. We also find that the temperature of the Local Group should
be higher than $2.3\times10^5$ K. To satisfy the observed \ion{O}{7} column
density, the gas distribution should have a rather flat
core with $r_{c} \geq 100$ kpc.

\subsection{Comments on a Hot Halo Gas Origin}

Strong \ion{O}{6} absorption along the sight line towards 3C~273 was
detected  with {\sl FUSE} between -100 and +100 $\rm km\ s^{-1}$
\citep{sem01a}. This absorption probably  occurs in the
interstellar medium of the Milky Way disk and halo. The column
density, $\rm \log N(O~VI) = 14.73\pm0.04$, is the second highest
O~{\sc vi} column recorded along any sight line through the Milky Way
halo (see \citealp{sav02}) . Absorption  features of lower ionization species are also
present at these velocities.  Most notably, the \ion{C}{4} 
$\lambda\lambda1548.195, 1550.770$ and \ion{N}{5} 
$\lambda\lambda1238.821,  1242.804$ lines are strong and track
the O~{\sc vi} absorption closely. Between -100 and +100 $\rm km\
s^{-1}$, N(\ion{O}{6})/N(\ion{C}{4}) and  N(\ion{O}{6})/N(\ion{N}{5})
are nearly constant, indicating that the  species arise in
the same gas along the sight line.

\ion{O}{6} absorption is also detected between +100 and +240 $\rm km\
s^{-1}$  in the form of a broad, shallow absorption wing extending
redward of the  primary  Galactic absorption feature.  This \ion{O}{6}
absorption wing has  $\rm \log N(O~VI) = 13.71$, or $\sim$ 10\% that of
the primary absorption feature  at lower velocities .  This wing is not observed in other species observable with
either {\sl HST} or {\sl FUSE}. The \ion{O}{6} absorption wing  has
been attributed to hot gas flowing out of the Galactic disk as part of
a  ``Galactic chimney'' or ``fountain'' in the Loop IV and North Polar
Spur regions  of the sky. Alternatively, the wing might be remnant
tidal debris from  interactions of the Milky Way and smaller Local
Group galaxies \citep{sem01a}. Similar \ion{O}{6}
absorption wings are seen in other directions in  this general region
of the sky \citep{sem02}.

It is possible to associate the  \ion{O}{7}
absorption detected by  {\sl Chandra} 
with these highly ionized metals detected by {\sl
FUSE} and {\sl HST}. Here, we briefly discuss  several scenarios:

{\it (1). The \ion{O}{7} is related to the primary \ion{O}{6}
feature.}
In this case, $\rm \log$ N(O~{\sc vi}) = 14.73, and  $\rm
\log$ [N(O~{\sc vii})/N(O~{\sc vi})]$\sim1.5$, assuming N(O~{\sc
vii})$\rm = 1.8\times10^{16}\ cm^{-2}$.  This is  within about a factor of
2 of the O~{\sc vii}/O~{\sc vi} ratio observed for the  PKS 2155-304
absorber at $z\approx0$ (\citealp{nic02,fan02}) and is consistent with the  idea that
the gas is radiatively cooling from a high temperature \citep{hns02}.  
This possibility is appealing since the  centroids of the
O~{\sc vi} and O~{\sc vii} absorption features are similar  ($\sim
6\pm10\ \rm km\ s^{-1}$  versus $-26\pm140\ \rm km\ s^{-1}$) and the
width of the resolved O~{\sc vi} line (FWHM $\sim$ 100 $\rm km\
s^{-1}$) is consistent with a broad O~{\sc vii} feature.  

However, this possibility also has drawbacks.  First, the amount of \ion{O}{8}
predicted by the \citet{hns02} cooling gas model would be $\sim$3
times the observed \ion{O}{7} column density. However, we did not detect any
\ion{O}{8} Ly$\alpha$ absorption line at the corresponded wavelength
in the  {\sl Chandra} spectrum. We estimate that the 4$\sigma$ upper
limit on the \ion{O}{8} Ly$\alpha$ equivalent width is $\sim 14.20\ \rm
m\AA$, or $\sim 10^{16}\ {\rm cm^{-2}}$ if the line is unsaturated. Second, the
amount of C~{\sc iv} predicted is a factor of $\sim$10 times less than
the O~{\sc vi} column density, which is a factor of 5 lower than
observed.  It is  possible that the C~{\sc iv} column could be
increased by including  photoionization from the cooling flow itself
\citep{bsh00} to bring this ratio closer to the observed
value.  The predicted N(O~{\sc vi})/N(N~{\sc v}) ratio of $\sim$20 is
about a factor of 3 higher than observed.  If there are  turbulent
mixing layers (TMLs) 
occurring between the hot gas and cooler ISM material (a likely
possibility), then the discrepancies in the O~{\sc vi}/C~{\sc iv}
ratio might be alleviated since the standard
TML models described by \citep{ssb93} predict considerably more C~{\sc iv} (but not more N~{\sc v}) than O~{\sc vi}.

{\it (2). The \ion{O}{7} is related to the high velocity \ion{O}{6} wing}.
In this
case, the \ion{O}{7} is associated only with the \ion{O}{6}
absorption ``wing'' and not with the main \ion{O}{6} absorption at 
lower velocities.  Although perhaps less likely than case (1), 
this may be plausible since the average velocity of the \ion{O}{6} absorption
wing ($\sim 150\rm\ km\ s^{-1}$) is still in reasonable agreement with
the \ion{O}{7} centroid.   In such a situation,
$\rm \log$ [N(O~{\sc vii})/N(O~{\sc vi})]$\sim2.5$.  In collisional
ionization equilibrium, this would imply a temperature of $>10^6$ K
(\citealp{smo76,sdo93}). In Figure~3 we
show the 90\% limit on  N(\ion{O}{7})/N(\ion{O}{6}) (the blue
region), which gives a tight constraint on the temperature of the
absorber:  $6.0 < \log T < 6.9$. A further constraint can be
achieved by considering the non-detection of \ion{O}{8} Ly$\alpha$
absorption. The 4$\sigma$ upper limit of column density requires 
$T\lesssim 10^{6.3}$ K; both N(\ion{O}{7})/N({\ion{O}{6}) and 
N(\ion{O}{8})/N({\ion{O}{7}) give about the
same constraint. 

Another possibility is that 
the \ion{O}{7} is not related the \ion{O}{6}. In this case, 
the temperature must be high enough ($> 10^6$ K) to prevent  
the production of \ion{O}{6}. The upper limit of the temperature can  
still be obtained by the N(\ion{O}{8})/N({\ion{O}{7}), which gives 
$T <  10^{6.3}$ K. This case is similar to that described in situation (2).

\subsection{Comments on Other Locations} 

Can other sources contribute to the observed \ion{O}{7} column
density? The  Local Bubble, a large volume of hot, rarefied gas
around the Sun, could  be a potential source. The hot Local Bubble 
gas has a temperature of $\sim10^6$ K \citep{sno98}, and contains highly 
ionized metal-line species (see \citealp{seim98}). 
However, given the
rather small size of the Local Bubble ($\lesssim 250$ pc),  even if 
the electron density is high ($\sim 0.005\rm\ cm^{-3}$; \citealp{slc99}) and
assuming all oxygen ions are in \ion{O}{7}, the overall  contribution
to the \ion{O}{7} column density is  $\lesssim 3\times
10^{15}\rm\ cm^{-3}$, or $\lesssim17$\% of the total \ion{O}{7} column density.
Another potential source is the hot gas in the Galactic thick disk / low halo
observed in \ion{O}{6} absorption \citep{sav02}. 
Given the high  latitude ($b=64.36^{\circ}$) of 3C~273, the path length 
through a scale height of this gas is $\sim 2$ kpc pc.  Assuming a
density of $\sim 0.001\rm\ cm^{-3}$, solar abundances, and unity ionization 
fraction (an upper bound), the overall contribution to  the \ion{O}{7} column 
density is still small ($\lesssim 30\%$) compared to the value observed.

The sight line to 3C~273 extends into the Galactic halo through the edges of Radio Loop I and IV, which have been attributed to supernova remnants (see, e.g., \citealp{eas95}). X-ray observations in this direction reveal a hot, X-ray emitting gas with a temperature of T $\sim 3\times 10^6$ K and emission measure ($n_e^2d$) $\sim (1-2) \times 10^{-2}\ \rm cm^{-6}pc$, where $d$ is the distance through Loops I and IV \citep{iwa80}. This gives a rough estimate of electron density $n_e \approx 0.01\ \rm cm^{-3}$. Using a path length of $< 350$ pc through Loops I and IV  \citep{ber73} and an \ion{O}{7} ionization fraction of 0.01 at $\rm T\sim 3\times 10^6$ K in collisional ionization equilibrium, the estimated column density for solar-abundance gas is $\lesssim 8\times10^{15}\ \rm cm^{-2}$, which could be as much as $\sim50$\% of the observed \ion{O}{7} column density if the loops really are this large.  (Iwan 1980 estimates a size 
$\lesssim 100$ pc.)   However, since the \ion{O}{7} ionization fraction raises quickly to 1 as the temperature approaches $10^6$ K, a small uncertainty on $T$ could cause a large variation in the predicted column density.

\section{Summary}

Our {\sl Chandra} observations indicate that a large amount of hot gas exists 
in the Local Group and/or in the Milky Way. The \ion{O}{7} column density in 
this direction is a factor of 2-2.5 higher than toward PKS~2155-304 (\citealp{nic02,fan02}) and a factor
of 2-5 higher than toward H\,1821+643 \citep{mwc02}, which combined with the 
strong X-ray emission from the North Polar Spur in the direction of 
3C~273, suggests that the absorption we see may well be from a combination 
of Galactic and Local Group sources. We constrain the temperature to $\rm 2.3\times 10^5 < T < 1.2\times 10^6$ K and find the baryon overdensity $\delta_b$ is $\sim 100$, assuming a standard $\beta$ model for the intragroup medium. The derived properties are 
consistent with those of the WHIM gas which are predicted from simulations. 
For the case of Galactic absorption, we have outlined several scenarios to 
explain the observed \ion{O}{7} absorption, which may be associated with
\ion{O}{6} observed at  UV wavelengths. We find that in both the Galactic or 
Local Group cases there is a
 strong indication that collisional ionization is the dominant ionization 
source, at least for the gas which produces \ion{O}{7} absorption. 
We expect future X-ray observations may provide more information on the 
distribution of \ion{O}{7} at $z \sim 0$ and its strength in different
 directions. Such observations would provide valuable insight needed 
to determine the the location of the hot gas. 

We obtained the data from the {\sl Chandra} Data Archive built by the {\sl Chandra} X-ray Center. We also thank the members of the MIT/CXC team for their support. This work is supported in part by contracts NAS 8-38249 and SAO SV1-61010.
KRS acknowledges financial support through NASA contract NAS5-32985
and Long Term Space Astrophysics grant NAG5-3485.

[Note: Recently, we learned from a conference presentation that \citet{ras02} reported the detection of a similar feature in the XMM spectrum of 3C 273, which confirms our detection.]

{}

\clearpage
\vbox{ 
\begin{center}

\begin{tabular}{llcc}
\multicolumn{2}{c}{~~~~~~~~Table 1: Line Fitting Parameters~~~~~~} \\
\hline \hline 
 &  O~{\sc vii}     & O~{\sc vii}     & O~{\sc viii}\\ 
 & He$\alpha$ & He$\beta$ & Ly$\alpha$ \\ \hline 
Wavelength (\AA) & $21.60\pm0.01$            & ... & ... \\
Rest Wavelength (\AA)$^{a}$ & 21.6019 & 18.6288 & 18.9689 \\
$cz$ (km~s$^{-1}$) & $-26 \pm 140$     & ... & ... \\ 
Line Width (\AA)$^{b}$ & $<0.02$                   & ... & ... \\ 
$\rm Line\ Flux^{c}$ & $4.15_{-0.90}^{+1.83}$           & ... & ... \\ 
$W_{\lambda}$ (m${\rm \AA}$)      & $28.4_{-6.2}^{+12.5}$ & $<11.9^d$ & $<14.2^d$\\ 
S/N & 6.4 & ... & ... \\ \hline
\end{tabular}

\parbox{3.5in}{
\vspace{0.1in} \small\baselineskip 9pt \footnotesize \indent
a. \citet{vvf96}.\\ b. 90\% upper limit of the line width $\sigma$.\\
c. Absorbed line flux, in units of $\rm
10^{-5}~photons~cm^{-2}s^{-1}$.\\ d. 4-$\sigma$ detection upper limit.
}
\end{center}
\normalsize \centerline{} }

\clearpage
\figcaption[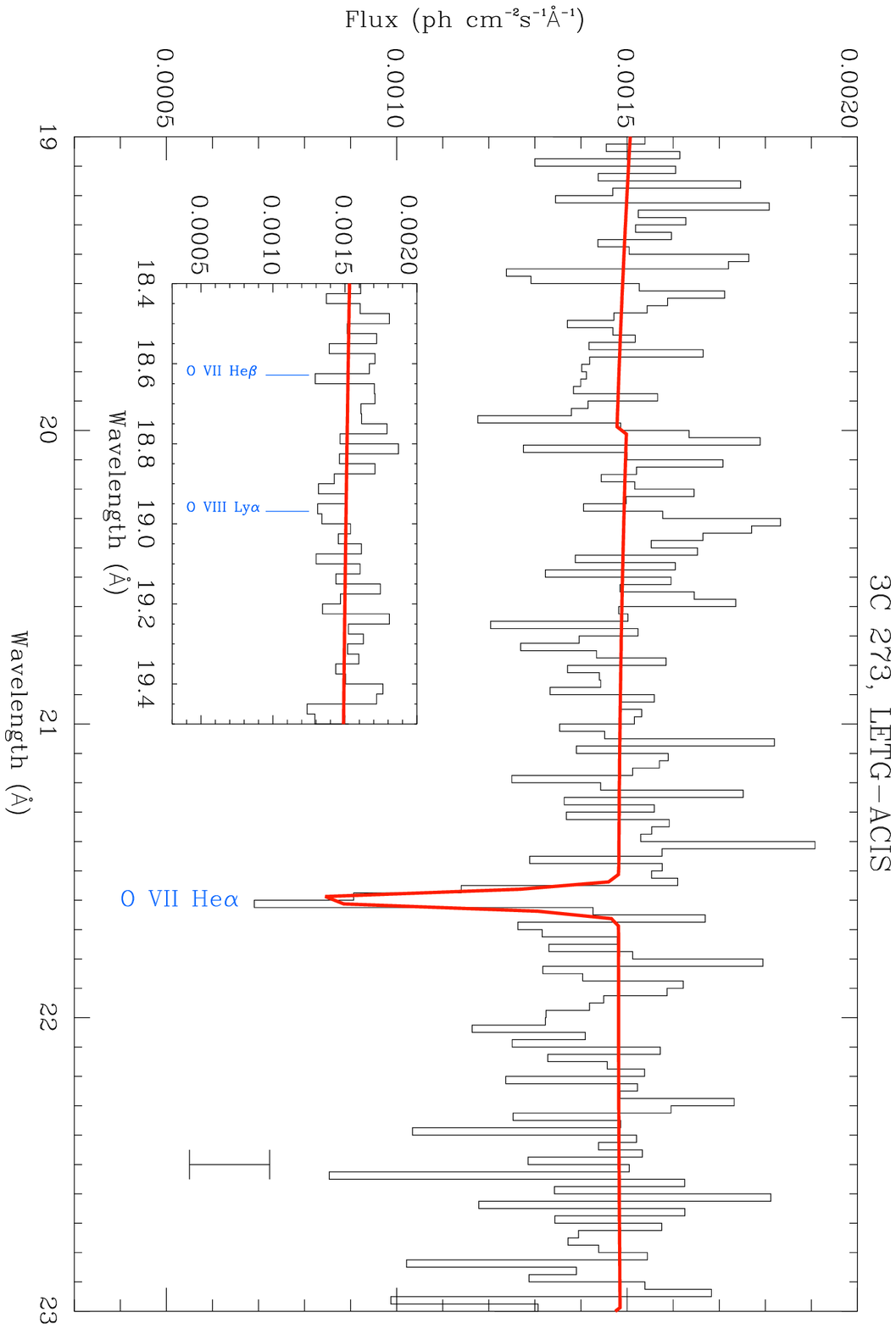]{The {\sl Chandra} LETG-ACIS spectrum of 3C273 between 19
and 23  $\rm\AA$. The red solid line is the fitted continuum (Galactic
absorbed power  law plus the polynomial) plus the Gaussian model. The
average $1\sigma$ error bar plotted on the right is based on
statistics only. The inset shows the  wavelength region where the rest-frame
\ion{O}{8} Ly$\alpha$ and \ion{O}{7} He$\beta$  are located. \label{fig1}}

\figcaption[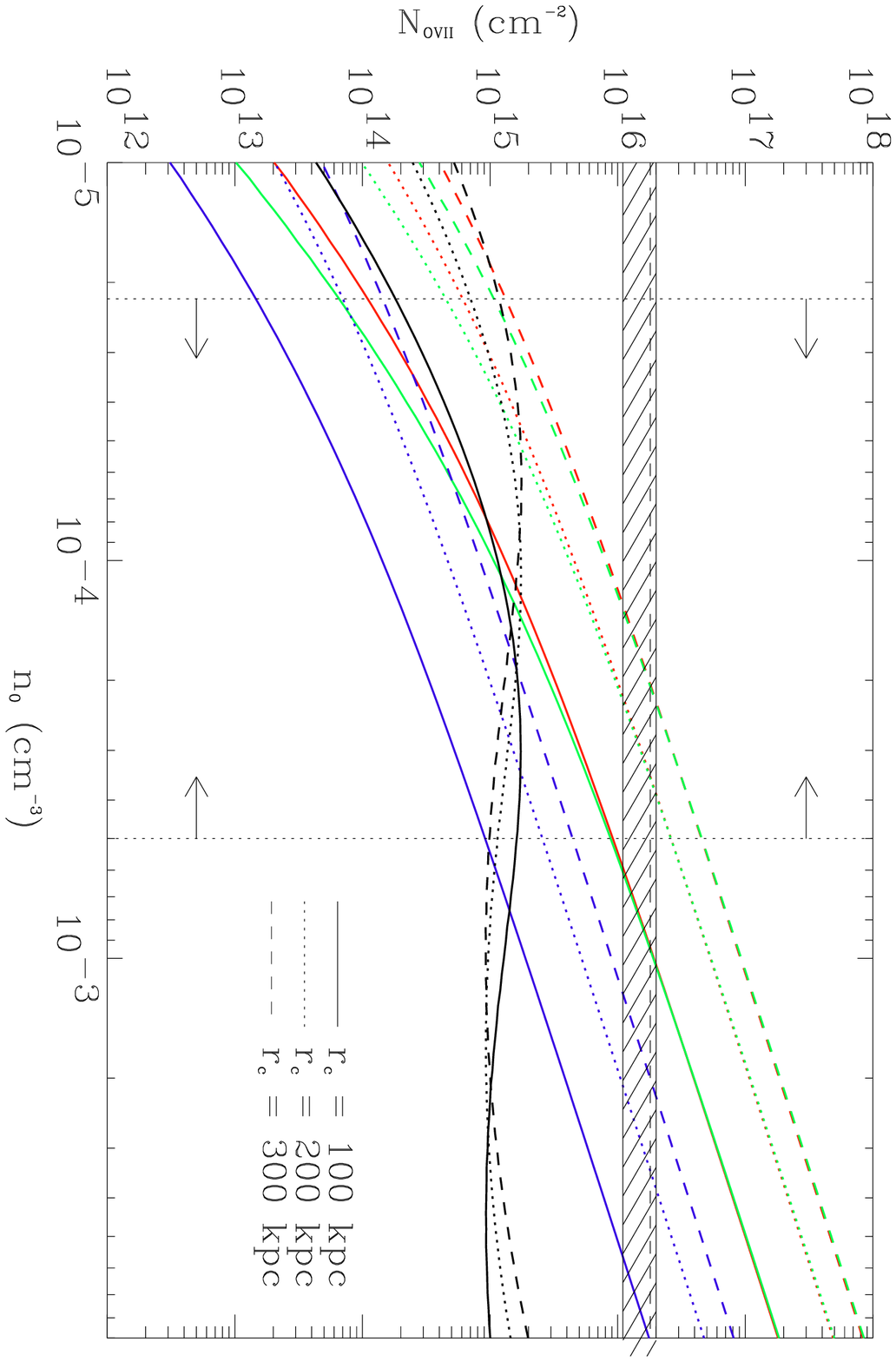]{O~{\sc vii} column density versus central baryon density. 
The four different colors denote four different Local Group temperatures 
[$T = 2.3\time10^5$ K (black), $5.8\times10^5$ K (red), $1.2\times10^6$ K (green) and $3.5\times10^6$ K (blue)]. 
The shadowed area indicates the 90\% confidence range of
the detected \ion{O}{7} column density. The two vertical dashed lines
indicate the central density lower limit  (from path length constraints)
and upper limit (from $\rm N_{O\,VII}$).\label{fig2}}

\figcaption[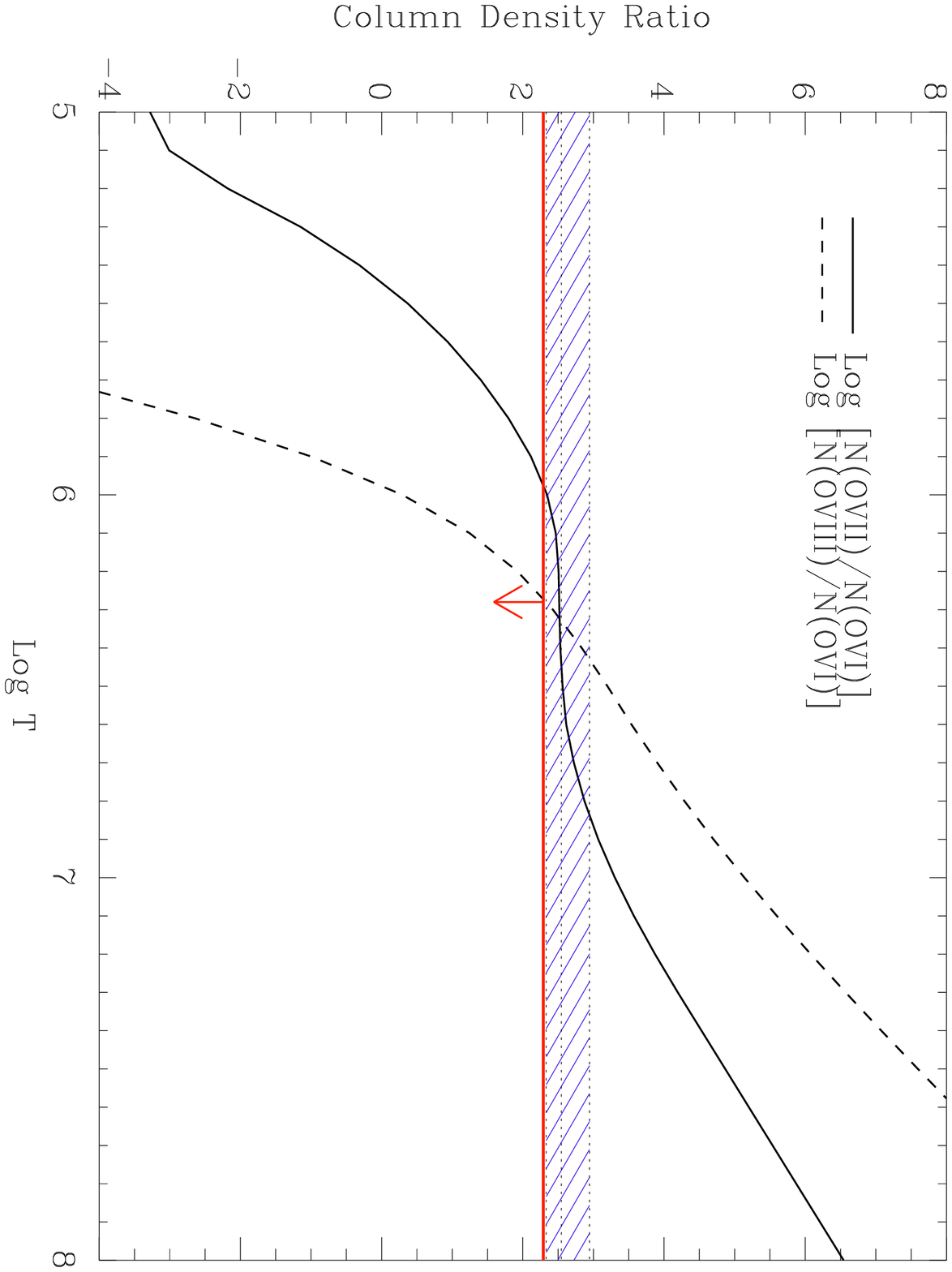]{The column density of \ion{O}{7} versus \ion{O}{6} (solid
line) and  \ion{O}{8} versus \ion{O}{6} (dashed line). The blue
region indicates the allowed range from the measured 
N(\ion{O}{7})/N(\ion{O}{6}) ratio. The red line is the lower limit of the
N(\ion{O}{8})/N(\ion{O}{6}) from the  non-detection of
\ion{O}{8} in the {\sl Chandra} spectra. \label{fig3}}

\clearpage
\begin{figure}
\epsscale{0.8}
\plotone{f1.eps}
\end{figure}

\clearpage
\begin{figure}
\epsscale{0.8}
\plotone{f2.eps}
\end{figure}

\clearpage
\begin{figure}
\epsscale{1.0}
\plotone{f3.eps}
\end{figure}

\end{document}